\documentclass[preprint,showpacs,amsmath,amssymb,aps,prd]{revtex4}
\usepackage{epsfig}
\begin{document}
\draft
\title{\mbox{}\\[10pt]
Phased Breaking of $\mu-\tau$ Symmetry and Leptogenesis}

\author{Y. H. Ahn$^{1,}$\footnote{E-mail:
        yhahn@cskim.yonsei.ac.kr},
        Sin Kyu Kang$^{2,}$\footnote{E-mail:
        skkang1@sogang.ac.kr},
        C. S. Kim$^{1,}$\footnote{E-mail:
        cskim@yonsei.ac.kr} and
        Jake Lee$^{2,}$\footnote{E-mail:
         jilee@cskim.yonsei.ac.kr}}

\address{$^{1}$  Department of Physics, Yonsei
        University, Seoul 120-749, Korea\\
        $^{2}$ Center for Quantum Spacetime, Sogang University,
        Seoul 121-742, Korea}



\date{\today}

\begin{abstract}

\noindent Non-vanishing $U_{e3}$ has been theoretically related to
a certain flavor symmetry breaking in the neutrino sector.
We propose a scenario to break the $\mu-\tau$ symmetry so as to
accommodate the non-vanishing $U_{e3}$.
Our scenario is constructed in the context of a seesaw model, and
the $\mu-\tau$ symmetry breaking is  achieved by
introducing a CP phase in the Dirac Yukawa matrix. We also show
how the deviation of $\theta_{23}$ from the maximal mixing and
non-vanishing $U_{e3}$ depend on the CP phase.
Neutrino mixings and the neutrino mass-squared differences are discussed,
and the amplitude in neutrinoless double beta decay $m_{ee}$
are also predicted. 
We found that a tiny breaking of the $\mu-\tau$ symmetry due
to mass splitting between two degenerate heavy Majorana neutrinos
on top of the Dirac CP phase can lead to successful leptogenesis.
We examine how
leptogenesis can be related with low energy neutrino measurement,
and show that our predictions for $U_{e3}$ and $m_{ee}$ can be
constrained by the current observation of baryon asymmetry.

\end{abstract}
\maketitle

\section{Introduction}

Thanks to the recent precise neutrino experiments, we now have
robust evidences for the neutrino oscillation. The present neutrino
experimental data \cite{atm,SK2002,SNO} indicate that the atmospheric
neutrino deficit points toward a maximal mixing between the tau and muon neutrinos,
however the solar neutrino deficit favors a not-so-maximal mixing
between the electron and muon neutrinos.
In addition, although we do not have yet any evidence
for the neutrino oscillation arisen from the 1st and 3rd generation flavor mixing,
there is a bound on the mixing element $U_{e3}$ from CHOOZ reactor
experiment, $|U_{e3}|<0.2$ \cite{chooz}.
Although neutrinos have gradually revealed their properties in various experiments
since the historic Super-Kamiokande confirmation of neutrino
oscillations \cite{atm}, properties related to the leptonic CP
violation are completely unknown yet. To understand in detail the neutrino
mixings observed in various oscillation experiments is one of the
most interesting puzzle in particle physics. The large value of
$\theta_{sol}$ and $\theta_{atm}$ may  be telling us about  some
new symmetries of leptons that are not present in the quark sector
and may provide a clue to understanding the nature of quark-lepton
physics beyond the standard model.

Recently, there have been some attempt to explain the maximal
mixing of the atmospheric neutrinos and very tiny value of the 3rd
mixing element $U_{e3}$  by introducing some approximate discrete
symmetries \cite{disc1,disc2} or the mass splitting among the
heavy Majorana neutrinos in the seesaw framework \cite{kangkim}.
In the basis where charged leptons are mass eigenstates, the
$\mu-\tau$ interchange symmetry has become useful in understanding
the maximal atmospheric neutrino mixing and the smallness of
$U_{e3}$ \cite{mu-tau,mu-tau0,mu-tau1,mu-tau2}. The mass
difference between the muon and the tau leptons, of course, breaks
this symmetry in the basis. So we expect this symmetry to
be an approximate one, and thus it must hold only for the neutrino
sector at low energy.

On the other hand, finding non-vanishing but small mixing element
$U_{e3}$ would be very interesting in the sense that the element
is closely related to leptonic CP violation \cite{ue3}. If future
neutrino experiments would measure the non-vanishing $U_{e3}$
\cite{jhf}, it may indicate from the afore-mentioned point of view
that the $\mu-\tau$ symmetry must be broken. Motivated by this
prospect, in this paper, we propose a scenario to break the
$\mu-\tau$ symmetry so as to accommodate the non-vanishing
$U_{e3}$. Our scenario is constructed in the context of a seesaw
model, and the symmetry breaking is first achieved by introducing
a CP phase in the Dirac Yukawa matrix.  Then, the resultant
effective light neutrino mass matrix generated through seesaw
mechanism reflects the $\mu-\tau$ symmetry breaking, which is
parameterized in terms of the CP phase. The $\mu-\tau$ symmetry
should be recovered in the limit of vanishing CP phase.

In fact, the breaking of the $\mu-\tau$ symmetry through a CP phase
has been also discussed in Ref. \cite{mohapatra1}, in which  the
authors have studied the breaking in the framework of
the effective light neutrino mass matrix.
In Ref. \cite{mu-tau2} the
breaking  has been also achieved  through the heavy Majorana neutrino mass
matrix with complex elements, which is completely different
from our scheme. We will study how neutrino mixings and the
neutrino mass-squared difference can be predicted and show how the
deviation of $\theta_{23}$ from the maximal mixing and
non-vanishing $U_{e3}$ depend on the CP phase in our scenario. The
prediction for the amplitude in neutrinoless double beta decay
will be discussed as well. However, it is turned out that the
Dirac CP phase introduced to break the $\mu-\tau$ symmetry does
not lead to successful leptogenesis, and thus we need a subsidiary
source for the lepton asymmetry. We will show that a tiny breaking
of the degeneracy between two heavy Majorana neutrinos, in
addition,
can lead to successful leptogenesis without much
affecting the predictions of the low energy neutrino observables.
We will also examine how leptogenesis can be related with low energy neutrino
measurement, and show that our predictions for $U_{e3}$ and
$m_{ee}$ can be constrained by the current observation of baryon
asymmetry.

\section{Neutrino sectors with $\mu-\tau$ symmetry}

To begin with, let us consider the Lagrangian of the lepton sector
from which the seesaw mechanism works,
\begin{eqnarray}
  {\cal L}_{m}=-Y_{\nu}^{ik}\overline{L_i}N_{k}\tilde{\phi}
              -Y_{l}^{i} \overline{L_i} l_{R_i} {\phi}
-\frac{1}{2}\overline{N_{k}}^{c} M_{N_k} N_{k}+h.c.,
\label{lagrangian}
\end{eqnarray}
where $i=e,\mu,\tau$ and $k=1,2,3$. $L_i,~ l_R,~ \phi,~ N$ are
$SU(2)$ lepton doublet fields, charged lepton singlet fields and
Higgs scalar, singlet heavy Majorana neutrino, respectively, and
$M_{N_k}$ denotes heavy Majorana neutrino masses. Here
we take a basis in which both charged lepton and singlet Majorana
neutrino mass matrices are real and diagonal.  The neutrino Dirac
Yukawa matrix with $\mu-\tau$ symmetry is given in the CP
conserving limit by
\begin{eqnarray}
Y_{\nu}= \left(
\begin{array}{ccc}
  a & b & b \\
  b & c & d \\
  b & d & c \\
\end{array}
\right),
\end{eqnarray}
where we assumed that $Y_{\nu}$ is symmetric. The Majorana mass
matrix $M_N$ with $\mu-\tau$ symmetry is given in the diagonal
form by $M_N={\rm Diag}[M_1,M_2,M_2]$. Note  the degeneracy of
$M_2=M_3$ from the presumed $\mu-\tau$ symmetry.  Later we will
discuss the effects of breaking the mass degeneracy, too. Then,
the effective light neutrino mass matrix generated through seesaw
mechanism reflects $\mu-\tau$ symmetry which in turn leads to
maximal mixing of the atmospheric neutrinos and vanishing mixing
angle $\theta_{13}$ in PMNS mixing matrix. In order to obtain
non-vanishing $\theta_{13}$, we have to break $\mu-\tau$ symmetry
appropriately. The $\mu-\tau$ symmetry breaking can generally be
achieved by imposing splittings between the same entries in the
mass matrices $Y_{\nu}$ and $M_N$. In this paper, however, we
break the symmetry by introducing a CP phase in the Dirac Yukawa
matrix $Y_{\nu}$ so that the same entries are distinguishable by
the phase. In principle, we can arbitrarily introduce CP phases in
the above Dirac Yukawa matrix to break the symmetry. However, we
note that any phases appearing in (2-3) sub-matrix of the
effective light neutrino mass matrix should be small to satisfy
the experimental result of $\Delta m_{sol}^2/\Delta m_{atm}^2$
\cite{mohapatra1}. In this regard, it is relevant to impose CP
phases in (2,1) and/or (3,1) entries of the Dirac Yukawa matrix
while keeping any entries in (2-3) sub-matrix of the effective
light neutrino mass matrix real. For simplicity, we take the (3,1)
entry to include a CP phase. Actually, either choice is turned out
to be completely equivalent. Incorporating a CP phase in (3,1)
entry of the Dirac mass matrix, the effective light neutrino mass
matrix through seesaw mechanism is given by
\begin{eqnarray} m_{\rm eff} &=& -\upsilon^{2}Y^{T}_{\nu}M^{-1}_{N}Y_{\nu} \nonumber \\
&=& -\upsilon^{2}{\left(\begin{array}{ccc}
 a & b &  b \\
 b & c & d \\
 be^{i\alpha} & d & c \end{array}\right)}^{T}{\left(\begin{array}{ccc}
 M_{1} & 0 &  0 \\
 0 & M_{2} & 0 \\
 0 & 0 & M_{2} \end{array}\right)}^{-1}\left(\begin{array}{ccc}
 a & b &  b \\
 b & c & d \\
 be^{i\alpha} & d & c \end{array}\right) \nonumber \\&=& -\upsilon^{2}\left(\begin{array}{ccc}
 \frac{a^{2}}{M_{1}}+\frac{b^{2}(1+e^{2i\alpha})}{M_{2}} & \frac{ab}{M_{1}}+\frac{b(c+de^{i\alpha})}{M_{2}} &  \frac{ab}{M_{1}}+\frac{b(d+ce^{i\alpha})}{M_{2}} \\
 \frac{ab}{M_{1}}+\frac{b(c+de^{i\alpha})}{M_{2}} & \frac{b^{2}}{M_{1}}+\frac{c^{2}+d^{2}}{M_{2}} & \frac{b^{2}}{M_{1}}+\frac{2cd}{M_{2}} \\
 \frac{ab}{M_{1}}+\frac{b(d+ce^{i\alpha})}{M_{2}} & \frac{b^{2}}{M_{1}}+\frac{2cd}{M_{2}} & \frac{b^{2}}{M_{1}}+\frac{c^{2}+d^{2}}{M_{2}} \end{array}\right).
\label{meff}
 \end{eqnarray}
As one can easily see,  the non-trivial value of  the CP phase $\alpha$
in the mass matrix $m_{\rm eff}$ breaks the $\mu-\tau$ symmetry.

\section{Neutrino mixing angles, a Dirac CP phase and neutrino mass spectrum}

First, we consider how the neutrino mixing angles and neutrino
mass spectrum can be predicted in our scenario. As can be expected
from the structure of the above $m_{\rm eff}$, in our model only the
normal hierarchical mass spectrum is allowed because the inverted
hierarchical case is achieved only when the off-diagonal elements
in heavy Majorana neutrino mass matrix are dominant \cite{king},
which is in contrast with the case of our model.
Furthermore, considering the
normal hierarchy and the maximality of the atmospheric neutrino
mixing, one can expect the following hierarchical structure among
the elements of the effective light neutrino mass matrix:
\begin{eqnarray}
|m_{\mu\tau,\mu\mu,\tau\tau}|\gg |m_{e\mu,e\tau}|\gg|m_{ee}|.
\label{cond1}
\end{eqnarray}
In terms of the light neutrino mass eigenvalues $m_i$, the above
condition (\ref{cond1}) leads to
$$
|m_{3}|\simeq|(m_{\rm eff})_{22}-(m_{\rm eff})_{23}|\gg |m_{2}|\simeq
|(m_{\rm eff})_{22}+(m_{\rm eff})_{23}|,
$$
and then we get the following
relations in terms of our parameters appeared in Eq. (\ref{meff}):
 \begin{eqnarray}
  && \frac{2|cd|}{M_{2}} \gg  \frac{b^{2}}{M_{1}}~,\quad cd < 0~, \label{rel_Ynu1} \\
  && |a| \ll |b| \ll  |c| \sim |d|~,\label{rel_Ynu2}
 \end{eqnarray}
where the degree of the hierarchy in Eq. (\ref{rel_Ynu2}) will
depend on that of the heavy Majorana masses $M_1$ and $M_2$.
Introducing new parameters from the ratios among the parameters given in Eq. (\ref{meff}),
\begin{eqnarray}
m_{0} \equiv \upsilon^{2}\frac{d^{2}}{M_{2}}~,~~ \eta \equiv \frac{M_1}{M_{2}}~,~~
\rho \equiv \frac{a}{d}~,~~ \omega \equiv \frac{b}{d}~,~~ \kappa \equiv \frac{c}{d}~, \label{newp}
\end{eqnarray}
 we can re-parameterize the
neutrino mass matrix $m_{\rm eff}$ as follows:
\begin{eqnarray} m_{\rm eff}=m_{0}{\left(\begin{array}{ccc}
 \frac{\rho^{2}}{\eta}+(e^{2i\alpha}+1)\omega^{2} & \frac{\rho\omega}{\eta}+(\kappa + e^{i\alpha})\omega &  \frac{\rho\omega}{\eta}+(1+\kappa e^{i\alpha})\omega \\
 \frac{\rho\omega}{\eta}+(\kappa + e^{i\alpha})\omega & \frac{\omega^{2}}{\eta}+1+\kappa^{2} & \frac{\omega^{2}}{\eta}+2\kappa \\
 \frac{\rho\omega}{\eta}+(1+\kappa e^{i\alpha})\omega & \frac{\omega^{2}}{\eta}+2\kappa  & \frac{\omega^{2}}{\eta}+1+\kappa^{2}  \end{array}\right)}.
\end{eqnarray}
Depending on the hierarchy of the heavy Majorana neutrino masses
$M_1$ and $M_2$, the relative sizes of the new parameters
consistent with the normal hierarchy of the neutrino mass spectrum
and the hierarchy of $\Delta m^2_{sol}$ and $\Delta m^2_{atm}$ can be
classified as follows:
\begin{eqnarray}
 &&\text{\bf Case 1}~ (M_{2}\gg M_{1}):  1\gg\eta\sim|\kappa|\eta\gg\omega \text{ or }
                   1\gg\omega\gg\eta|\kappa|\sim\eta\gg\eta\omega, \nonumber \\
 &&\text{\bf Case 2}~ (M_{2}\simeq M_{1}): 1\sim\eta\sim|\kappa|\eta\gg\omega, \label{cases}\\
 &&\text{\bf Case 3}~ (M_{2}\ll M_{1}):\eta\gg|\kappa|\gg\omega\nonumber.
\end{eqnarray}
Note that $\kappa$ is negative as can be seen in the relation
(\ref{rel_Ynu1}), and  the  quantity $\rho/\eta$ is
very small compared to the other ones in $m_{\rm eff}$. For numerical
purpose, we consider the case of $\rho=0$ without a loss of
generality. Then the neutrino mass matrix $m_{\rm eff}$ is simplified
as
\begin{eqnarray} m_{\rm eff}=m_{0}{\left(\begin{array}{ccc}
 (e^{2i\alpha}+1)\omega^{2} & (\kappa + e^{i\alpha})\omega &  (1+\kappa e^{i\alpha})\omega \\
 (\kappa + e^{i\alpha})\omega & \frac{\omega^{2}}{\eta}+1+\kappa^{2} & \frac{\omega^{2}}{\eta}+2\kappa \\
 (1+\kappa e^{i\alpha})\omega & \frac{\omega^{2}}{\eta}+2\kappa  & \frac{\omega^{2}}{\eta}+1+\kappa^{2}
 \end{array}\right)}
 \equiv
{\left(\begin{array}{ccc}
 \tilde{p} & \tilde{q} &  \tilde{q}^\prime \\
 \tilde{q} & r & s \\
 \tilde{q}^\prime & s  & r
 \end{array}\right)},
 \label{matrix}
\end{eqnarray}
where the complex variables are distinguished by the tilde.
This neutrino mass matrix is diagonalized by the PMNS mixing
matrix $U_{\rm PMNS}$,
$U^{T}_{\rm PMNS}m_{\rm eff}U_{\rm PMNS}={\rm Diag}[m_1,m_2,m_3]$,
where $m_{i}~ (i=1,2,3)$ indicates
the mass eigenvalues of light Majorana neutrinos. But, we
diagonalize the hermitian matrix $ m^{\dag}_{\rm eff}m_{\rm eff}$ \cite{harrison, aizawa} instead,
so that we can easily obtain the mixing angles and
phases appeared in $U_{\rm PMNS}$ in terms of the parameters
appeared in Eq. (\ref{matrix}),
 \begin{eqnarray}
  m^{\dag}_{\rm eff}m_{\rm eff}=U_{\rm PMNS}~{\rm Diag}
  (m^{2}_{1},m^{2}_{2},m^{2}_{3})~U^{\dag}_{\rm PMNS}
  \equiv  {{\left(\begin{array}{ccc}
  A & \tilde{B} &  \tilde{C} \\
  \tilde{B}^{\ast} & D & \tilde{E} \\
  \tilde{C}^\ast & \tilde{E}^{\ast} & D \end{array}\right)}},
 \end{eqnarray}
 where
\begin{eqnarray}
  && A=|\tilde{p}|^2 + |\tilde{q}|^2 + |\tilde{q}^\prime|^2,\quad
    \tilde{B}= \tilde{p}^\ast\tilde{q}+\tilde{q}^\ast r+\tilde{q}^{\prime\ast}s,\quad
  \tilde{C}= \tilde{p}^\ast\tilde{q}^\prime+\tilde{q}^\ast s+\tilde{q}^{\prime\ast}r, \nonumber \\
  && D=|\tilde{q}|^2 + r^2 + s^2,\quad\quad ~
   \tilde{E}= \tilde{q}^\ast\tilde{q}^\prime+2rs.  {}
 \end{eqnarray}
 Here we note that $A$ and $D$ are real.
 Then, the straightforward calculation with the standard parametrization
 of $U_{\rm PMNS}$ leads to the
 expressions for the masses and mixing parameters \cite{aizawa}:
\begin{eqnarray}
m_{1,2}^2 &=& \frac{\lambda_1+\lambda_2}{2}\mp\frac{c_{23}{\rm Re}(\tilde{B})-s_{23}
                       {\rm Re}(\tilde{C})}{2s_{12}c_{12}c_{13}},\quad
m_3^2=\frac{c_{13}^2\lambda_3-s_{13}^2A}{c_{13}^2-s_{13}^2}, \\
\tan\theta_{23} &=&  \frac{{\rm Im}(\tilde{B})}{{\rm Im}(\tilde{C})},~~
\tan 2\theta_{12}=2\frac{c_{23}{\rm Re}(\tilde{B})-s_{23}{\rm Re}(\tilde{C})}{c_{13}(\lambda_2-\lambda_1)}, ~~
\tan 2\theta_{13}=2\frac{|s_{23}\tilde{B}+c_{23}\tilde{C}|}{\lambda_3-A}, \\
\tan\delta &=& -\frac{1}{s_{23}}\frac{{\rm Im}(\tilde{B})}{s_{23}{\rm Re}(\tilde{B})+c_{23}{\rm Re}(\tilde{C})},
\end{eqnarray}
with
\begin{eqnarray}
\lambda_1=c_{13}^2A-2s_{13}c_{13}|s_{23}\tilde{B}+c_{23}\tilde{C}|+s_{13}^2\lambda_3,\quad
\lambda_{2,3}=D\mp2s_{23}c_{23}{\rm Re}(\tilde{E}).
\end{eqnarray}
As can be seen from Eqs. (10-16), three neutrino masses, three
mixing angles and a CP phase are presented in terms of five
independent parameters $m_0, \omega, \kappa, \eta, \alpha$.  At
present, we have five experimental results, which are taken as
inputs in our numerical analysis given  at $3\sigma$  by
\cite{maltoni},
\begin{eqnarray}
&&28.7^\circ < \theta_{12} < 38.1^\circ,\quad 35.7^\circ <
\theta_{23} < 55.6^\circ,\quad
0^\circ < \theta_{13} < 13.1^\circ,\nonumber\\
&&7.1<\Delta m^2_{21}[10^{-5}{\rm eV}^2]<8.9,\quad 1.4<\Delta
m^2_{31}[10^{-3}{\rm eV}^2]<3.3\label{exp_bound}.
\end{eqnarray}
Imposing the current experimental results on neutrino masses and mixings
into the above relations (13-16) and  scanning
all the parameter space $\{ m_0, \omega, \kappa, \eta, \alpha\}$,
we investigate how those parameters are constrained and estimate
possible predictions for other phenomena such as neutrino-less
double beta decay and leptonic CP violation.


\begin{figure}[tb]
\vspace*{-5.0cm}
\hspace*{-6cm}
\begin{minipage}[t]{6.0cm}
\epsfig{figure=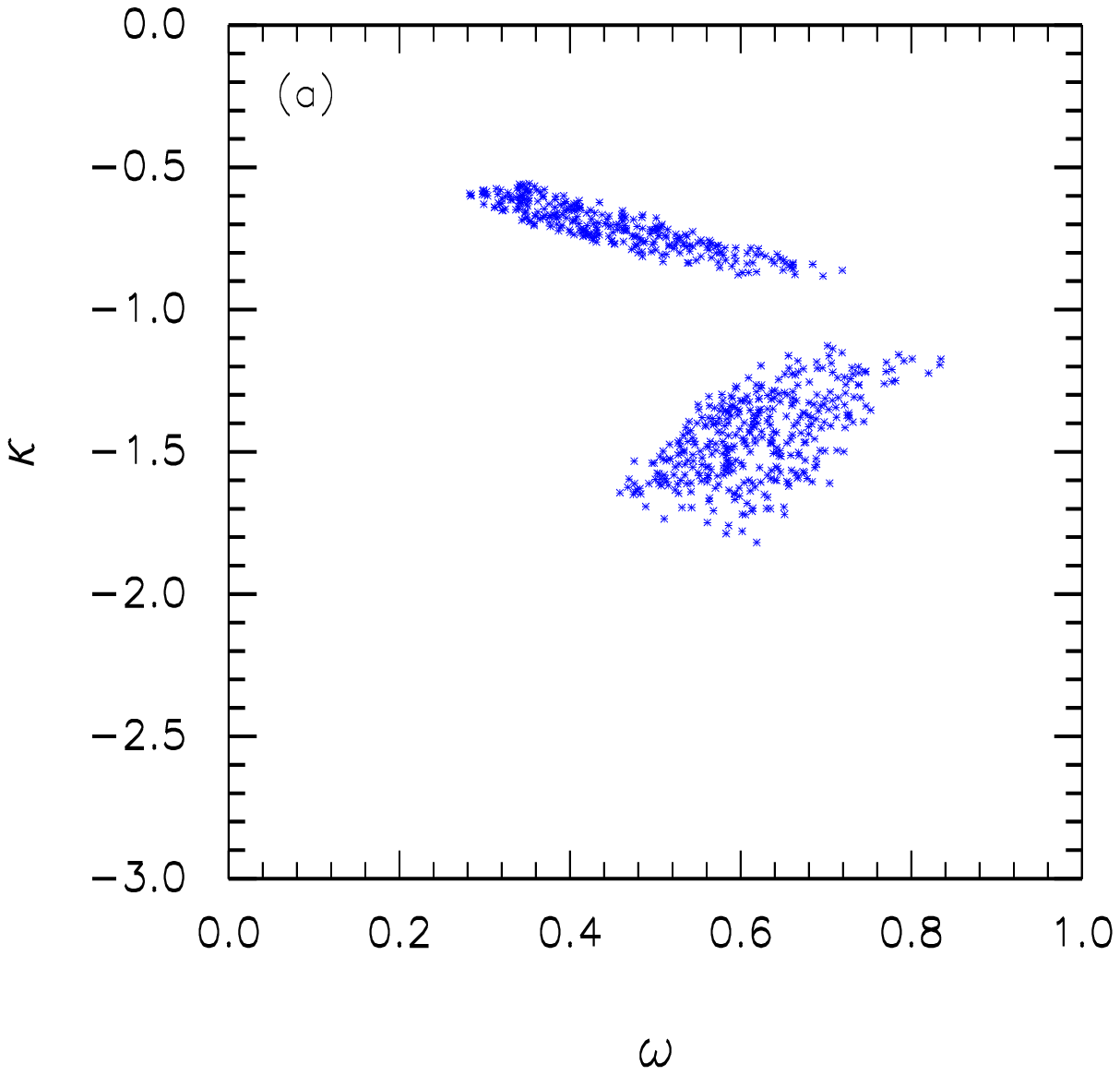,width=13cm,angle=0}
\end{minipage}
\hspace*{2.0cm}
\begin{minipage}[t]{6.0cm}
\epsfig{figure=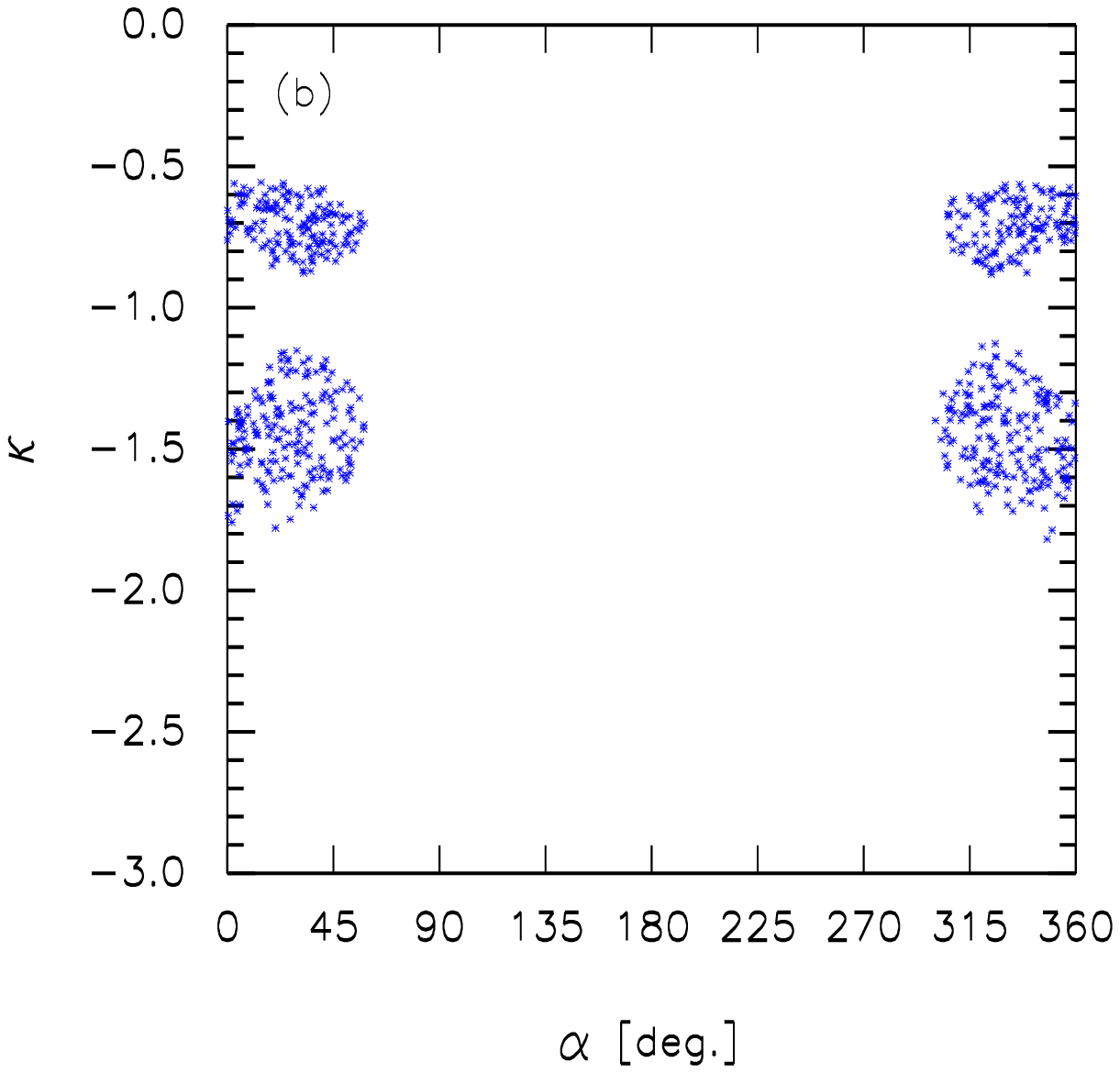,width=13cm,angle=0}
\end{minipage}
\vspace*{-5.5cm}
\caption{\label{Fig1}Allowed parameter regions by the $3\sigma$ experimental constraints
in Eq.~(\ref{exp_bound}) of Case 2 ($M_{2}\simeq M_{1}$) in Eq.~(\ref{cases}). Here we take
$\eta=1$.}
\end{figure}

\begin{figure}[tb]
\vspace*{-5.0cm}
\hspace*{-6cm}
\begin{minipage}[t]{6.0cm}
\epsfig{figure=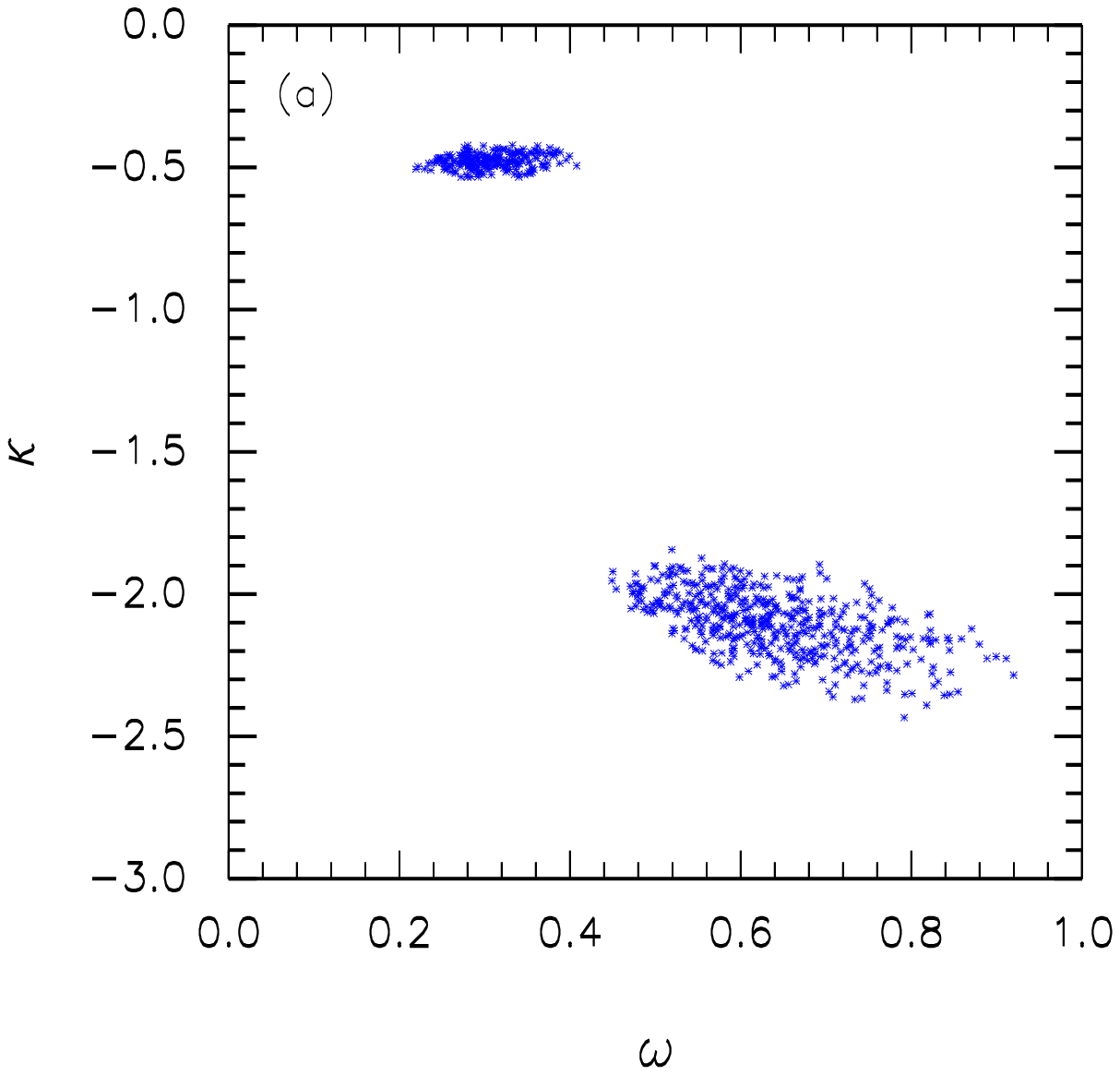,width=13cm,angle=0}
\end{minipage}
\hspace*{2.0cm}
\begin{minipage}[t]{6.0cm}
\epsfig{figure=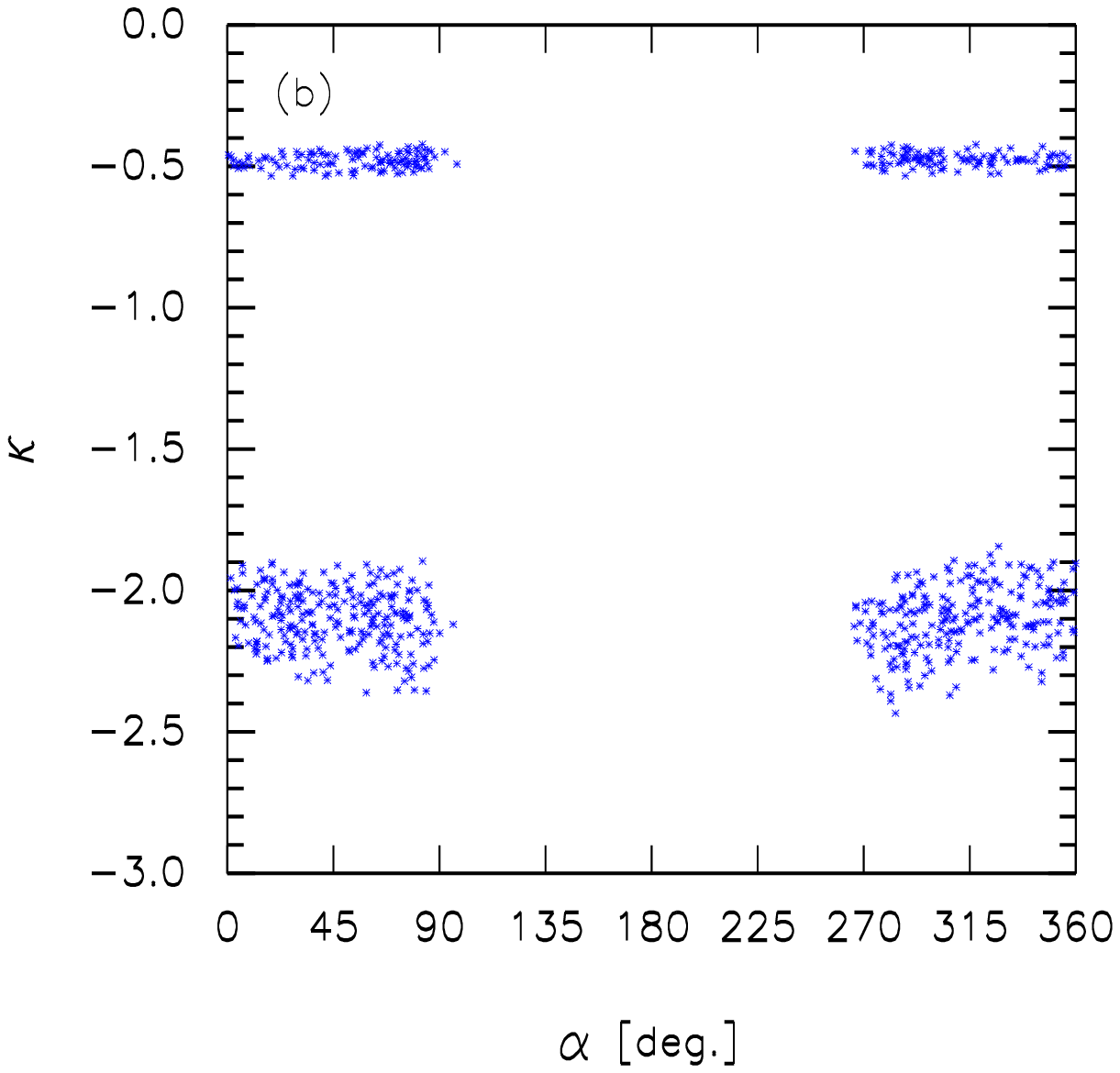,width=13cm,angle=0}
\end{minipage}
\vspace*{-5.5cm}
\caption{\label{Fig2}Allowed parameter regions by the $3\sigma$ experimental constraints
in Eq.~(\ref{exp_bound}) for Case 3 ($M_{1}\gg M_{2}$) in Eq.~(\ref{cases}). Here we take
$\eta=1000$.}
\end{figure}

Let us discuss the numerical results focusing on the three cases given in (\ref{cases}).
As a result of the numerical analysis concerned with the mixing angle $\theta_{12}$,
we found that the Case 1 ($M_2 \gg M_1$) is ruled out
mainly because we get a very small $\theta_{12}$ for all the parameter space
in this case. So, we will focus on Case 2 and Case 3.
In Fig.~\ref{Fig1}, we show the parameter regions constrained by the experimental
results given in Eq. (\ref{exp_bound}) for $\eta=1$. The two figures present how the parameter
$\kappa$
can be constrained depending on the parameter $\omega$ and the phase $\alpha$, respectively.
The allowed range of $m_0$ is turned out to be $10^{-3} \sim 10^{-2}$ eV.
In Fig.~\ref{Fig2}, we show the same constrained parameter regions for Case 3.
Here we fix $\eta=1000$, however, we found that the dependence of $\eta$ on the allowed regions
of the other parameters is very weak as long as
$\eta\ge 10$.

We note that the most severe constraint for the parameters comes from
the solar mixing angle $\theta_{12}$.
To see how the solar mixing angle constrain the parameters, it is useful to consider
the approximate form of $\theta_{12}$ in the limit of maximal $\theta_{23}$ and tiny $\theta_{13}$,
which is given by
 \begin{eqnarray}
  \tan2\theta_{12}&\simeq& \frac{2\sqrt{2}(1+\kappa)\omega\cos^{2}\frac{\alpha}{2}\{(1+\kappa)^{2}
  +2\frac{\omega^{2}}{\eta}+2\omega^{2}\cos\alpha\}}
  {(1+\kappa)^{4}+(\kappa-1)^{2}\omega^{2}(\cos\alpha-1)+4(1+\kappa)^{2}\frac{\omega^{2}}
  {\eta}+4\frac{\omega^{4}}{\eta^{2}}}.
 \end{eqnarray}
Based on the above expression, let us discuss the predictions of the mixing angle $\theta_{12}$ case
by case classified in (\ref{cases}).
\begin{itemize}
\item For Case 2, the solar mixing angle is further approximated:
       \begin{eqnarray}
       \tan2\theta_{12}&\simeq& \frac{2\sqrt{2}(1+\kappa)\omega\cos^{2}\frac{\alpha}{2}\{(1+\kappa)^{2}
       +2\omega^{2}(1+\cos\alpha)\}}
       {(1+\kappa)^{4}+(\kappa-1)^{2}\omega^{2}(\cos\alpha-1)+4\omega^{2}\{(1+\kappa)^{2}
       +\omega^{2}\}}. \label{solmix}
        \end{eqnarray}

As can be seen in Fig.~\ref{Fig1}, the current experimental values
of $\theta_{12}$ are achieved only when the condition
$|1+\kappa|\sim|\omega|$ is satisfied. Due to this condition, it
appears that two allowed regions are separated in Fig.~\ref{Fig1}.
\end{itemize}


\begin{itemize}
\item For Case 3, the expression for $\theta_{12}$ is further simplified as
       \begin{eqnarray}
        \tan2\theta_{12}&\simeq&
        \frac{2\sqrt{2}(1+\kappa)\omega\cos^{2}\frac{\alpha}{2}\{(1+\kappa)^{2}+2\omega^{2}\cos\alpha\}}
        {(1+\kappa)^{4}+(\kappa-1)^{2}\omega^{2}(\cos\alpha-1)}. \label{solmix}
        \end{eqnarray}

In this case we found that only the parameter regions leading to $|1+\kappa|\ge|\omega|$ are allowed
by the result from the solar neutrino experiments.
\end{itemize}

\begin{itemize}
\item For vanishing phase $\alpha=0$, we can easily
see that  $\theta_{23}=\frac{\pi}{4}$, $\theta_{13}=0$ and
$\tan2\theta_{12}=\frac{2\sqrt{2}\omega(1+\kappa)}{(1+\kappa)^{2}+2\omega^{2}(\frac{1}{\eta}-1)}$,
which can be consistent with the result of the solar neutrino
experiments. However, $\alpha=\pi$ is not allowed because it would
result in $\theta_{12}=0$ as can be seen from Eq. (\ref{solmix}).
\end{itemize}

\begin{figure}[tb]
\vspace*{-5.0cm}
\hspace*{-6cm}
\begin{minipage}[t]{6.0cm}
\epsfig{figure=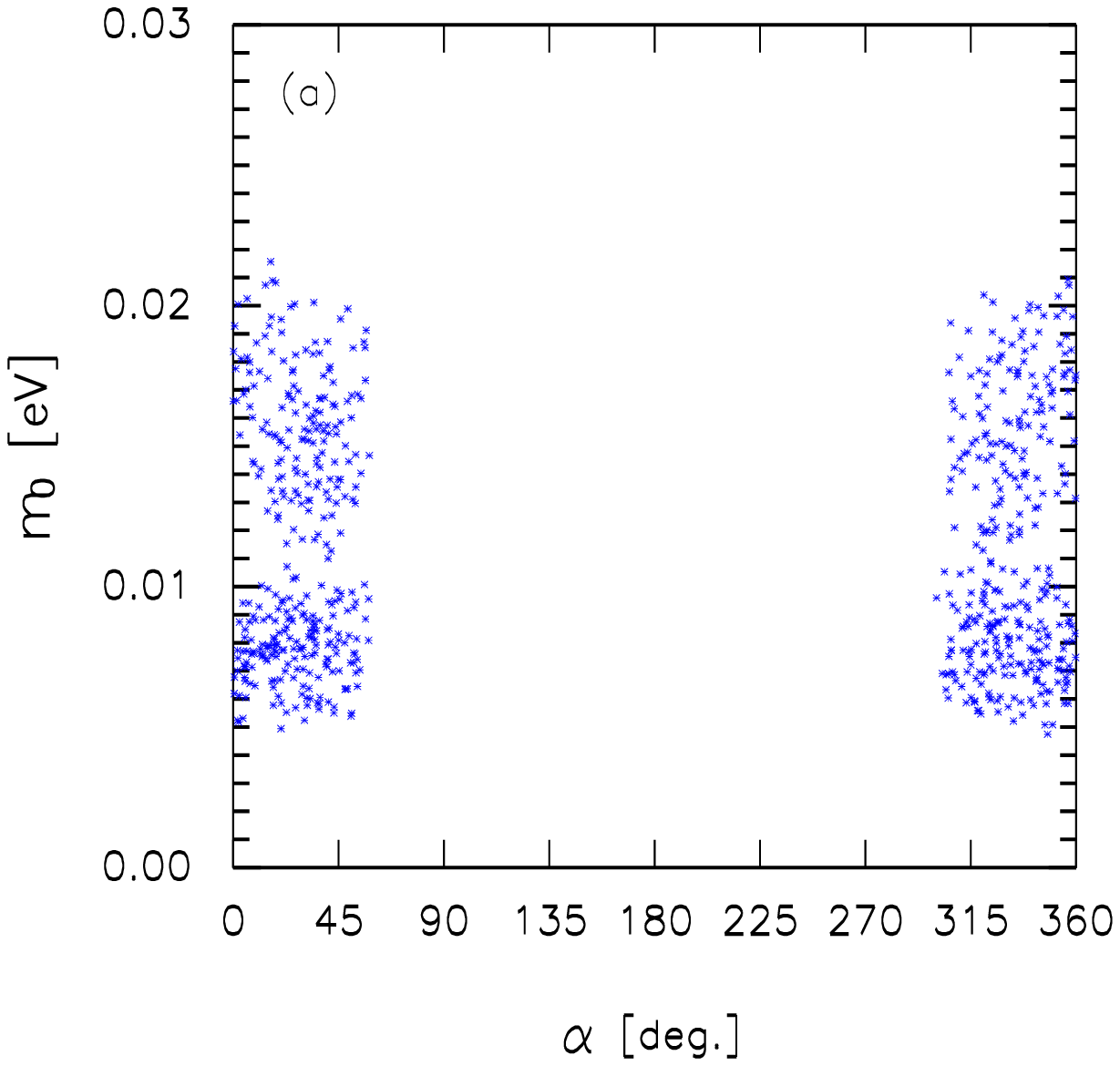,width=13cm,angle=0}
\end{minipage}
\hspace*{2.0cm}
\begin{minipage}[t]{6.0cm}
\epsfig{figure=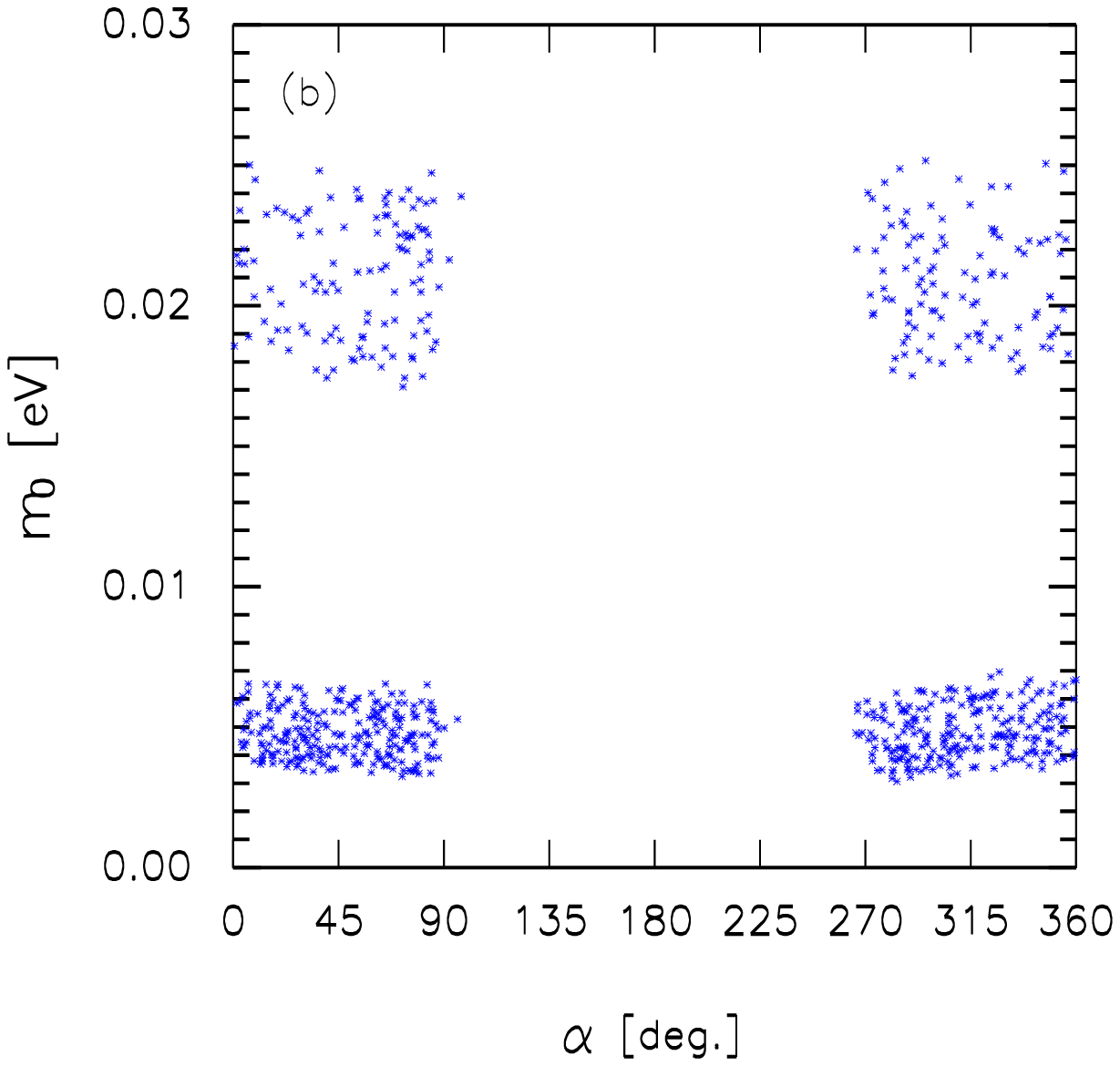,width=13cm,angle=0}
\end{minipage}
\vspace*{-5.5cm}
\caption{\label{Fig2-1} Predictions of $m_0$ over allowed parameter regions
by the $3\sigma$ experimental constraints
in Eq.~(\ref{exp_bound}) for (a) Case 2 and (b) Case 3 in Eq.~(\ref{cases}).
}
\end{figure}

In Fig.~\ref{Fig2-1}, we present the prediction of the parameter $m_0$, which determines
the overall mass scale of the light neutrinos, as a function of $\alpha$ for $\eta=1~(1000)$.
Combining the allowed regions for the parameters $\kappa$ and $\omega$ shown in
Fig.~\ref{Fig1} and Fig.~\ref{Fig2},
$m_0$ in our model is turned out to be of order $10^{-3} \sim 10^{-2}$ eV, which is around
the atmospheric scale $m_0 \sim \sqrt{\Delta m^2_{atm}}/2$ as expected from the fact that
our scenario is relevant for the normal hierarchical light neutrino mass spectrum.


\begin{figure}[tb]
\vspace*{-5.0cm}
\hspace*{-6cm}
\begin{minipage}[t]{6.0cm}
\epsfig{figure=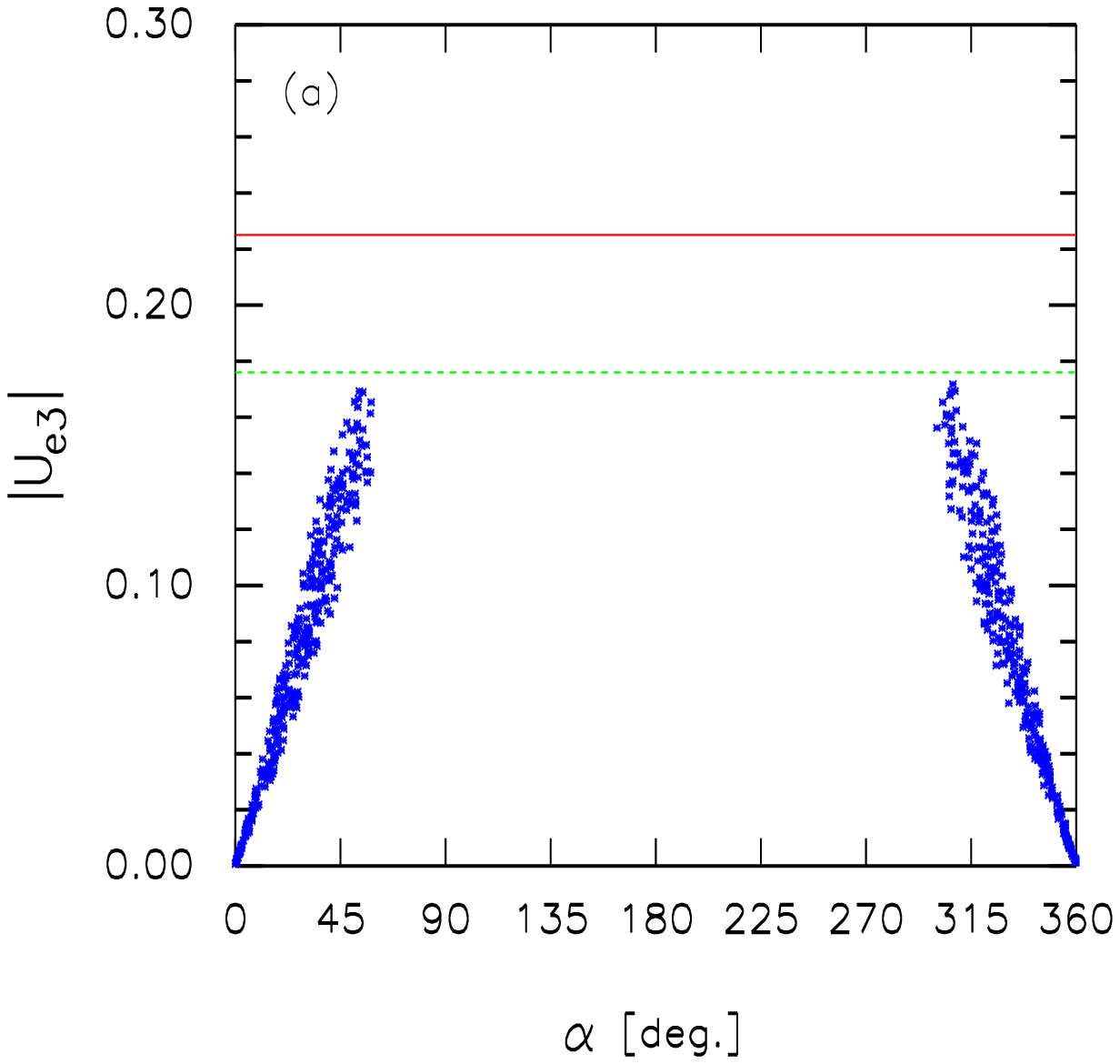,width=13cm,angle=0}
\end{minipage}
\hspace*{2.0cm}
\begin{minipage}[t]{6.0cm}
\epsfig{figure=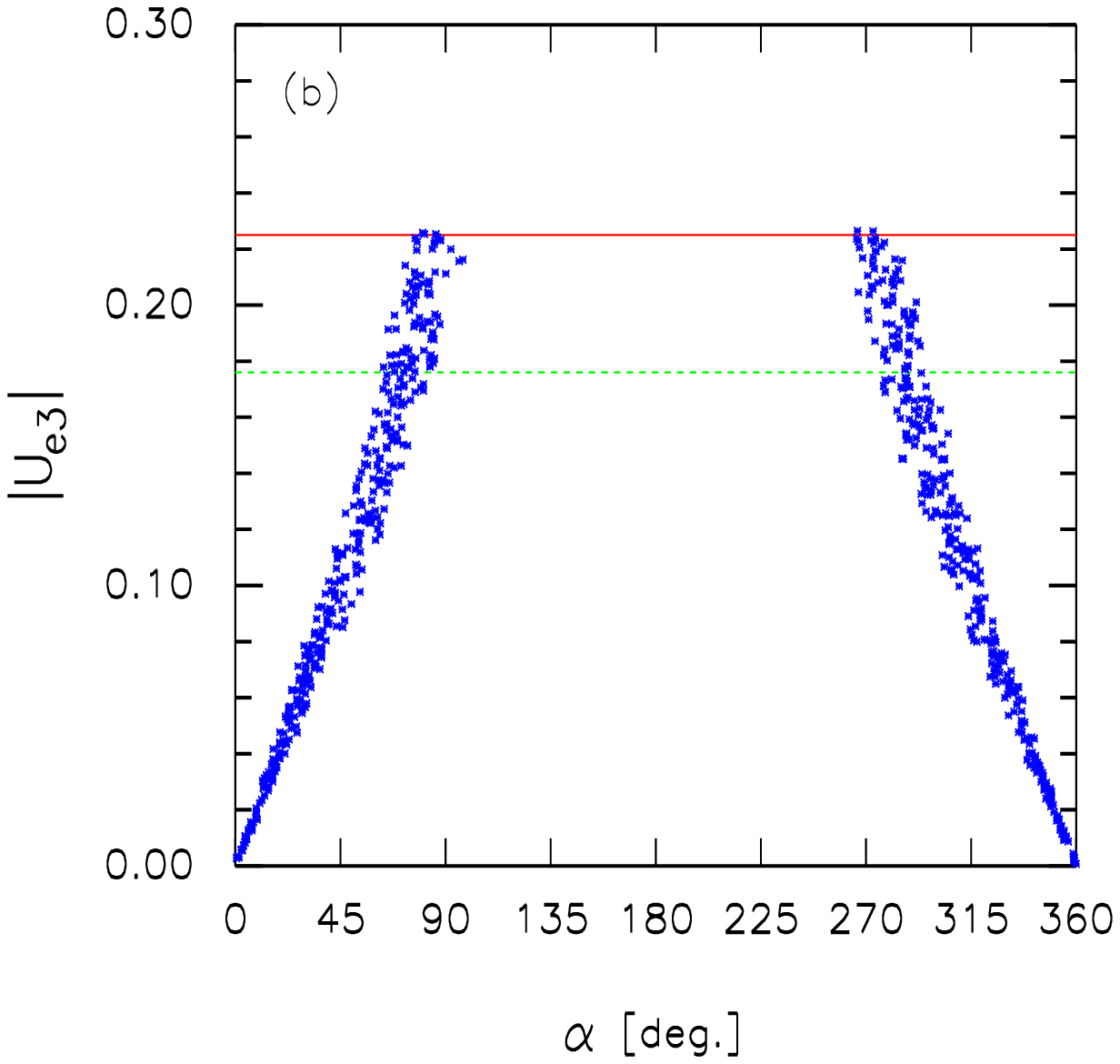,width=13cm,angle=0}
\end{minipage}
\vspace*{-5.5cm}
\caption{\label{Fig3} Predictions of $|U_{e3}|$ over allowed parameter regions
by the $3\sigma$ experimental constraints
in Eq.~(\ref{exp_bound}) for (a) Case 2 and (b) Case 3 in Eq.~(\ref{cases}).
The horizontal solid (dotted) lines are $3\sigma$ ($2\sigma$) upper bounds, respectively.}
\end{figure}

Next, we consider how our scenario predicts the sizes of
$\theta_{13}$ and Dirac phase $\delta$. In Fig.~\ref{Fig3}, we show
the predictions of $|U_{e3}|$ as a function of the phase $\alpha$
for Case 2 and Case 3. The horizontal solid (dotted) lines
correspond to the experimental bound on $\theta_{13}$ from CHOOZ
experiment at $3\sigma~ (2\sigma)$ C.L., respectively. For Case 2,
$|U_{e3}|$ is predicted well below the current bound. Thus, the
current experimental bound on $\theta_{13}$ does not constrain the
parameter space. Contrary to Case 2, Fig. 4(b) shows that the
experimental bound on $\theta_{13}$ can constrain the parameter
regions for Case 3.  We remark that we have cut the points above the
$3\sigma$ bound in Fig. 4(b). In the parameter regions leading to
the best-fit points of the neutrino mixing angles, we obtain the
following approximate expressions for $U_{e3}$ and the  phase angle
$\delta$ of $U_{\rm PMNS}$:
 \begin{eqnarray}
  |U_{e3}|&\simeq & \frac{\omega}{|\kappa-1|}\sqrt{1-\cos\alpha}, \\
   \sin\delta &\simeq &-\cos\frac{\alpha}{2}.
 \end{eqnarray}
Interestingly enough, we see that the non-vanishing $|U_{e3}|$
depends on the non-trivial value of $\alpha$ and it is also
related with the phase $\delta$. These approximate
expressions are the same as those given in Ref. \cite{mohapatra1}.

The atmospheric mixing angle $\theta_{23}$ is also deviated from the
maximal mixing. The deviation is approximately given by
\begin{eqnarray}
 \theta_{23}-\frac{\pi}{4}\simeq\frac{(1-\kappa^{2})}{4\kappa(1+\kappa^{2})}\omega^{2}\sin^{2}\frac{\alpha}{2}.
\end{eqnarray}
We see from the above expression that the atmospheric neutrino
mixing goes to maximal for $\alpha=0$, and the parameter regions
consistent with the solar neutrino and CHOOZ experiments indicate
that the deviation from maximality of the atmospheric mixing angle
so small that it is well below the experimental limit of
$\theta_{23}$.

\begin{figure}[tb]
\vspace*{-5.0cm}
\hspace*{-6cm}
\begin{minipage}[t]{6.0cm}
\epsfig{figure=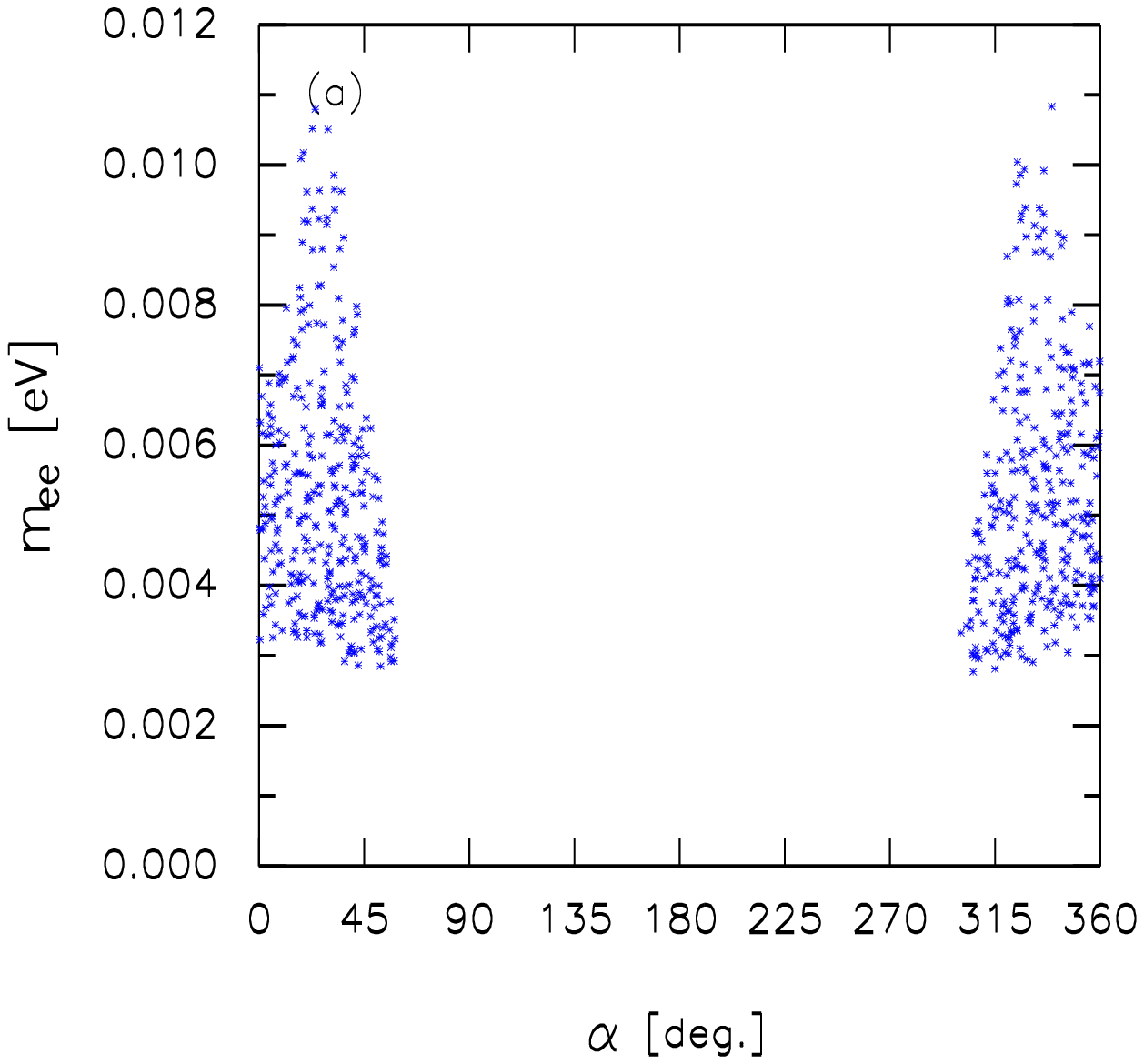,width=13cm,angle=0}
\end{minipage}
\hspace*{2.0cm}
\begin{minipage}[t]{6.0cm}
\epsfig{figure=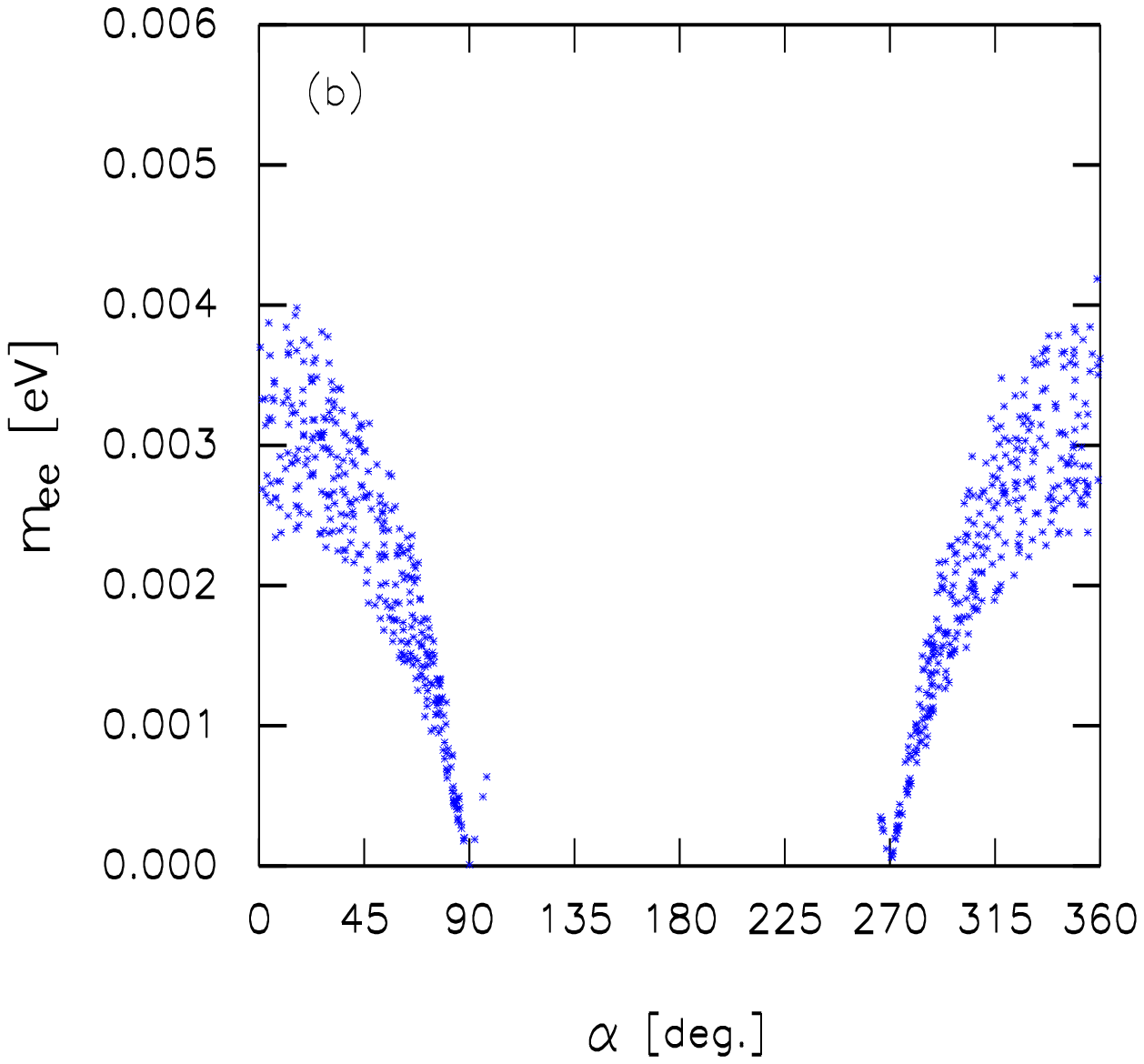,width=13cm,angle=0}
\end{minipage}
\vspace*{-5.5cm}
\caption{\label{Fig4}Predictions for the effective mass $m_{ee}$
for the neutrinoless double $\beta$-decay over allowed parameter regions
by the $3\sigma$ experimental constraints
in Eq.~(\ref{exp_bound}) for (a) Case 2 and (b) Case 3 in Eq.~(\ref{cases}).}
\end{figure}

Now let us consider the neutrinoless double beta decay which is
related with the absolute value of the $ee$-element of the light
neutrino mass matrix and is approximately given in our scenario  by
 \begin{eqnarray}
  |\langle m_{ee}\rangle|&\simeq& |m_{0}(e^{2i\alpha}+1)\omega^{2}|\nonumber\\
                       &=& m_{0}\omega^{2}\sqrt{2(1+\cos2\alpha)}.
 \end{eqnarray}
As can be seen in the above equation, $m_{ee}$ vanishes for $\alpha=\pi/2$ or $3\pi/2$
in our model.
In Fig.~\ref{Fig4}, we show the predictions for $m_{ee}$ as a function of the phase $\alpha$.
Case 2 predicts larger $m_{ee}$ than Case 3, and
even gives lower limits, $m_{ee} > 0.003$ eV.


\section{Leptogenesis}

We now consider how leptogenesis can work out in our scenario.
For the decay of a heavy Majorana neutrino $N_i$,
the CP asymmetry generated through the interference between tree
and one-loop diagrams is
given by \cite{lepto,lepto2}:
 \begin{eqnarray}
  \text{CP Asymmetry: }\varepsilon_{i}=\frac{\Gamma(N_{i}\rightarrow l\varphi)
  -\Gamma(N_{i}\rightarrow \overline{l}\varphi^{\dag})}{\Gamma(N_{i}\rightarrow l\varphi)
  +\Gamma(N_{i}\rightarrow\overline{l}\varphi^{\dag})}
  =\frac{1}{8\pi}\sum_{k\neq i}
  \frac{{\rm Im}[Y_{\nu}Y^{\dag}_{\nu}]^{2}_{ik}}{(Y_{\nu}Y^{\dag}_{\nu})_{ii}}\tilde{f}
  (\frac{M^{2}_{k}}{M^{2}_{i}}),\nonumber
 \end{eqnarray}
where $M_i$ denotes the heavy Majorana neutrino masses and the loop function $\tilde{f}(x_{i})$
containing vertex and self-energy corrections is
 \begin{eqnarray}
  \tilde{f}(x)=\sqrt{x}\left((1+x)\ln\frac{x}{1+x}+\frac{2-x}{1-x}\right).
 \end{eqnarray}
We note that the asymmetry $\varepsilon_1$ due to the decay
of the heavy Majorana neutrino $N_1$ vanishes because the CP phase
does not show up in the relevant terms due to $\rho=0$:
 \begin{eqnarray}
{\rm Im}[Y_{\nu}Y^{\dag}_{\nu}]^{2}_{i1}={\rm
Im}[Y_{\nu}Y^{\dag}_{\nu}]^{2}_{1i}=0. \label{rho}
 \end{eqnarray}
In fact, non-vanishing but small $\rho~(\equiv a/d)$, whose
size is constrained by neutrino data, can
lead to non-zero $\varepsilon_1$. However, the numerical value of
lepton asymmetry generated from non-vanishing
$[Y_{\nu}Y^{\dag}_{\nu}]^{2}_{i1}$ is still too small for successful
leptogenesis.

Since there are no contributions of $N_1$ to $\varepsilon_{2(3)}$
due to Eq. (\ref{rho}), the lepton asymmetry can be generated only
in case that the degeneracy between $N_2$ and $N_3$ is broken. And
quasi-degeneracy is still desirable because it does not much
affect the results for low energy neutrino observables obtained in
sec. III. We find that even a tiny mass splitting between $N_2$
and $N_3$ on top of the $\mu-\tau$ symmetry breaking through the
Dirac CP phase can lead to successful leptogenesis. We expect that
lepton asymmetry is resonantly enhanced in our case
\cite{resonant1, resonant2}. For convenience, we introduce a
parameter $\delta_N$ representing the degree of degeneracy as
follows:
 \begin{eqnarray}
  \delta_{N} \equiv \frac{M_{3}-M_{2}}{M_{2}}.
 \end{eqnarray}
Since $\delta_N$ is taken to be very small,
the CP asymmetries $\varepsilon_{2(3)}$ are approximately given by
 \begin{eqnarray}
  \varepsilon_{2} &\simeq &\frac{-1}{16\pi(Y_{\nu}Y^{\dag}_{\nu})_{22}}\left\{\frac{{\rm Im}[(Y_{\nu}Y^{\dag}_{\nu})^{2}_{23}]}
  {\delta_{N}}\right\}, \nonumber \\
  \varepsilon_{3} &\simeq &\frac{1}{16\pi(Y_{\nu}Y^{\dag}_{\nu})_{33}}\left\{\frac{{\rm Im}[(Y_{\nu}Y^{\dag}_{\nu})^{2}_{32}]}
  {\delta_{N}}\right\},
   \label{eps}
 \end{eqnarray}
 where we have used $\tilde{f}[(1+\delta_{N})^{\pm2}]\simeq\mp(1/2\delta_{N})$ for
$\delta_N \ll 1$ \cite{resonant2}. The relevant Yukawa terms in
our scenario are as follows,
 \begin{eqnarray}
  (Y_{\nu}Y^{\dag}_{\nu})_{22(33)}&=& d^{2}(1+\kappa^{2}+\omega^{2}),\nonumber\\
  {\rm Im}[(Y_{\nu}Y^{\dag}_{\nu})^{2}_{23}]&=& -2d^{4}\omega^{2}(2\kappa+\omega^{2}\cos\alpha)\sin\alpha, \nonumber\\
  {\rm Im}[(Y_{\nu}Y^{\dag}_{\nu})^{2}_{32}]&=& 2d^{4}\omega^{2}(2\kappa+\omega^{2}\cos\alpha)\sin\alpha,
 \end{eqnarray}
 and then the resulting CP asymmetries are given by
 \begin{eqnarray}\label{epsilon}
  \varepsilon_{2}&= & \varepsilon_{3}\simeq\frac{1}{8\pi}\frac{d^{2}\omega^{2}(2\kappa+\omega^{2}\cos\alpha)\sin\alpha}
  {\delta_{N}(1+\kappa^{2}+\omega^{2})}.
 \end{eqnarray}


\begin{figure}[tb]
\vspace*{-5.0cm}
\hspace*{-6cm}
\begin{minipage}[t]{6.0cm}
\epsfig{figure=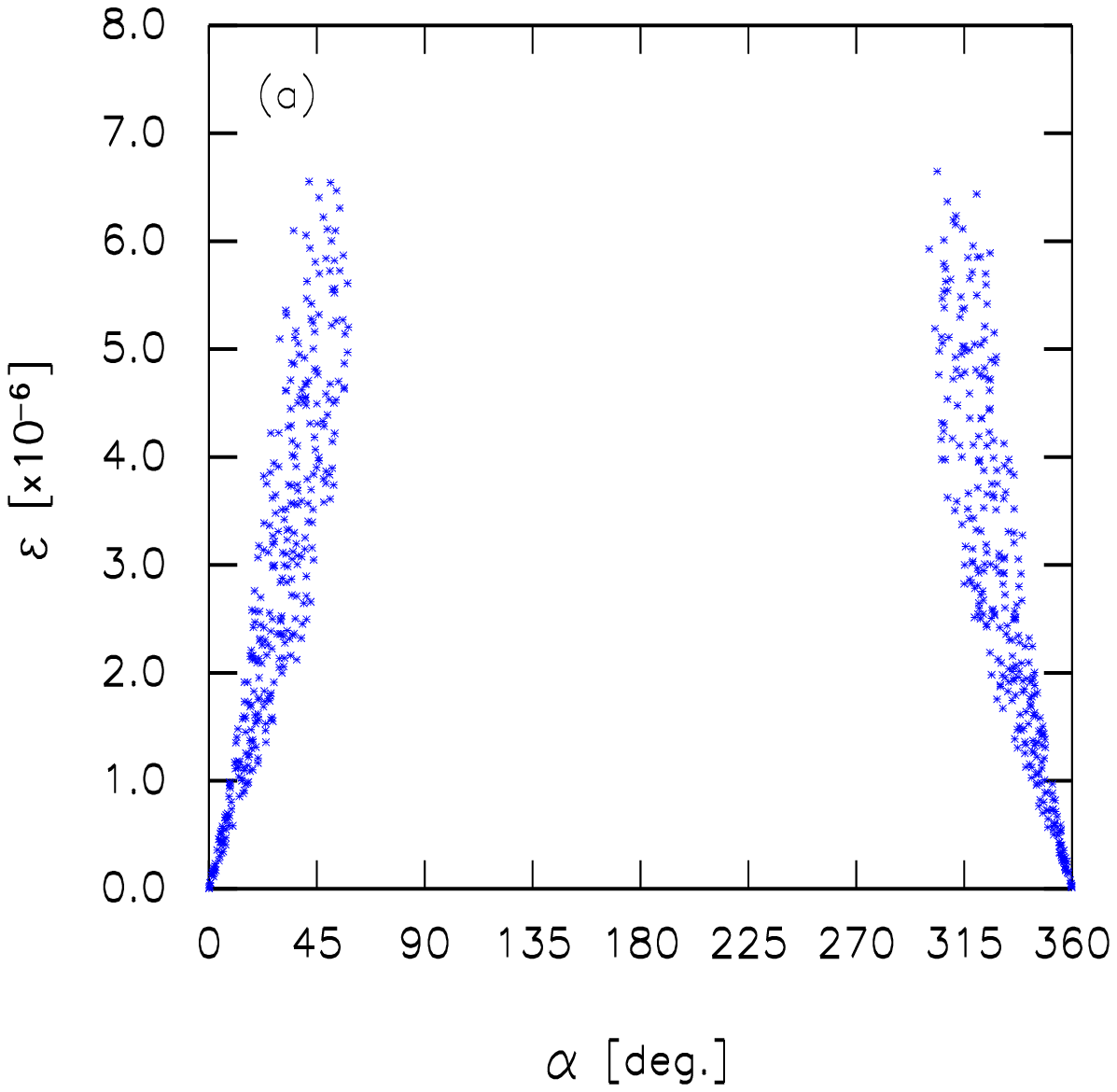,width=13cm,angle=0}
\end{minipage}
\hspace*{2.0cm}
\begin{minipage}[t]{6.0cm}
\epsfig{figure=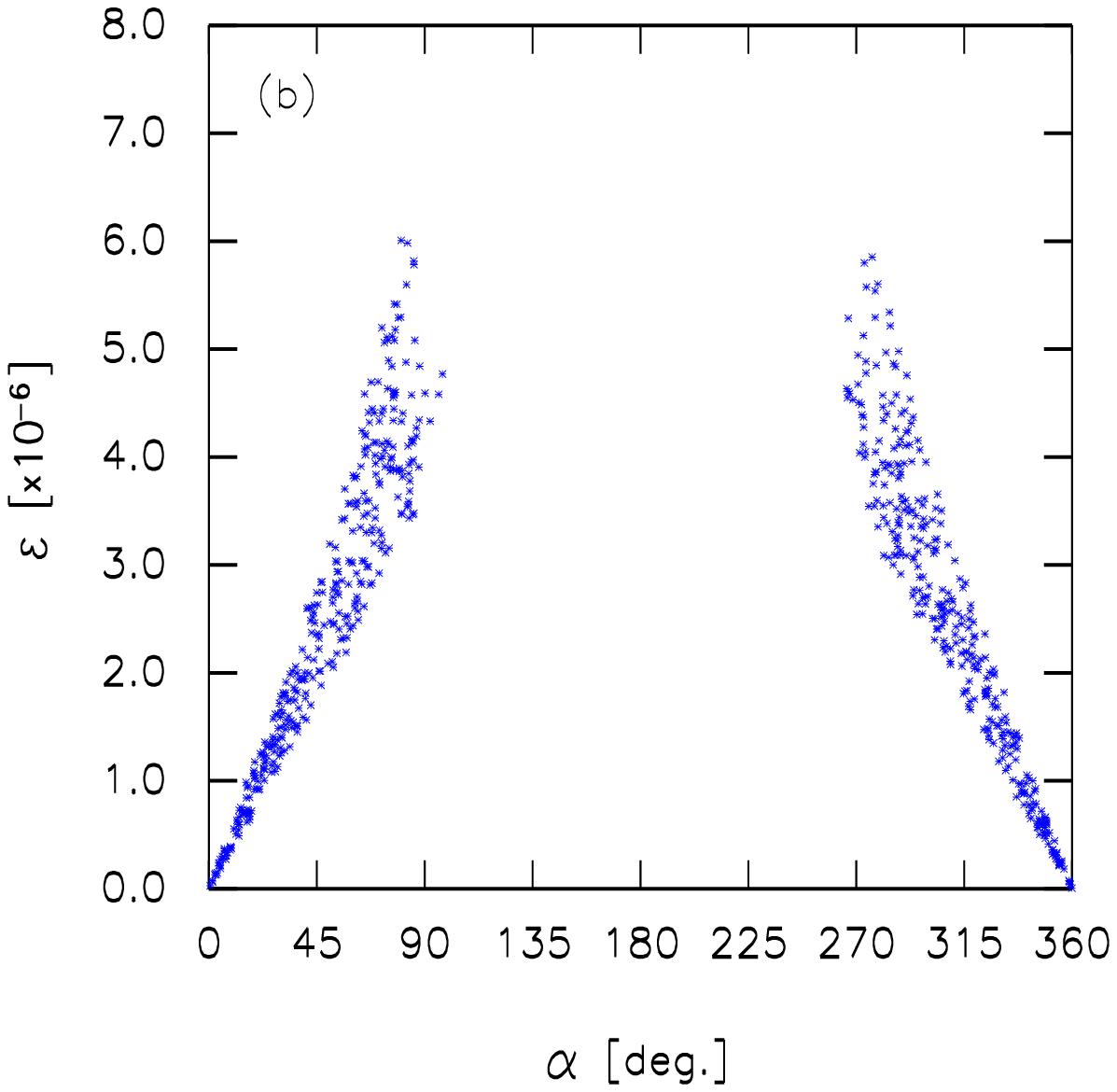,width=13cm,angle=0}
\end{minipage}
\vspace*{-5.5cm}
\caption{\label{Fig5}Predictions for the CP asymmetry $\epsilon$
for resonant leptogenesis over allowed parameter regions
by the $3\sigma$ experimental constraints
in Eq.~(\ref{exp_bound}) for (a) Case 2 and (b) Case 3 in Eq.~(\ref{cases}).
The dependence on $d$(or $M_2$) and $\delta_N$ is proportional to each other, that is,
this figure corresponds to $M_2=10^{10}$(GeV) with $\delta_N=10^{-2}$
or $M_2=10^{8}$(GeV) with $\delta_N=10^{-4}$, etc.}
\end{figure}
In this expression, the values of the parameters $\omega, \kappa, \alpha$ are determined
from the analysis described in sec. III, whereas $\delta_N$ and $d$ are arbitrary.
However, since $d^2=m_0 M_2/v^2$ as defined in Eq. (\ref{newp}), the value of $d$ depends on
the magnitude of $M_2$ in the case that $m_0$ is determined as before.
Thus, in our numerical analysis, we take $M_2$ and $\delta_N$ as input in the estimation
of lepton asymmetry. Here, we note that although $\delta_N$ and $M_2$ are taken to be
independent parameters in our analysis, the predictions of the lepton asymmetry $\varepsilon_{2(3)}$
depends only on the quantity $M_2/\delta_N$.
In Fig.~\ref{Fig5}, we show the predictions of the total lepton asymmetry
for the specific values of $\delta_N$ and $M_2$.
It is likely from Eq.~(\ref{epsilon}) that one could arbitrarily
enhance the asymmetry by lowering $\delta_N$. However, the value
of $\delta_N$ is constrained from the validity of the
perturbation. In order for the perturbative approach to be valid,
the tree-level decay width $\Gamma_i$ must be much smaller than
the mass difference:
 \begin{eqnarray}
 \Gamma_i=\frac{[Y_{\nu}Y^{\dag}_{\nu}]_{ii}}{8\pi}M_{i}
  \ll M_3-M_2=\delta_N M_2,\quad i=2,3.
 \end{eqnarray}
 Numerically, our model requires
$\delta_N \gg 10^{-7}$ for $M_2=10^{10}\;{\rm GeV}$,
 and so the maximum degeneracy or the minimum $\delta_N$
 in our scenario could be
 \begin{eqnarray}
\delta^{\rm min}_N \sim 10^{-6} \times \left(\frac{M_2}{10^{10}\;{\rm GeV}}\right).
 \end{eqnarray}

Now, let us study how we can achieve successful baryon asymmetry in our model.
Actually, the resulting baryon-to-photon ratio can be
estimated as
 \begin{eqnarray}
  \eta_{B}\simeq 10^{-2}\sum_{i}\varepsilon_{i}\cdot\kappa_{i}
 \end{eqnarray}
where the efficiency factor $\kappa_{i}$ describe the washout of
the produced lepton asymmetry $\varepsilon_{i}$.
The efficiency in generating the resultant baryon asymmetry
is usually controlled by the parameter defined as
 \begin{eqnarray}
 K_i \equiv \frac{\Gamma_i}{H} = \frac{\tilde{m}_i}{m_\ast},
 \end{eqnarray}
where $\Gamma_i$ is the tree-level decay width of $N_i$
and $H$ is the Hubble constant.
Here, the so-called effective neutrino mass, $\widetilde{m}_i$ is
\begin{eqnarray}
\widetilde{m}_{i}=\frac{[m_{D}m^{\dag}_{D}]_{ii}}{M_{i}},
 \end{eqnarray}
and
${m_\ast}$ is defined as
\begin{eqnarray}
  m_{\ast} &=& \frac{16
  \pi^{\frac{5}{2}}}{3\sqrt{5}}g^{\frac{1}{2}}_{\ast}\frac{\upsilon^{2}}{M_{\rm Planck}}\simeq
  1.08\times10^{-3}\;{\rm eV},
 \end{eqnarray}
 where we adopted $M_{\rm Planck}=1.22\times 10^{19}\;{\rm GeV}$ and
 the effective number of degrees of freedom
$g_{\ast} \simeq g_{\ast{\rm SM}}=106.75$.
Although most analyses on baryogenesis via leptogenesis
conservatively consider
$K_i < 1$, much larger values of $K_i$, even larger than $10^3$, can
be tolerated \cite{pilaftsis}.
Using the expression of $\widetilde{m}_i$ in terms of our parameters defined above,
\begin{eqnarray}
\widetilde{m}_i=m_0(1+\kappa^2+\omega^2),
 \end{eqnarray}
we find that our scenario resides in so-called {\it strong washout regime} with
\begin{eqnarray}
20 \lesssim K_i \lesssim 30.
 \end{eqnarray}
So, for numerical calculations, we adopt an approximate expression
of the efficiency factor applicable for large $K_i$ \cite{nielsen},
\begin{eqnarray}
\kappa_i \approx \frac{0.3}{K_i(\ln K_i)^{0.6}}.
 \end{eqnarray}


\begin{figure}[tb]
\vspace*{-5.0cm}
\hspace*{-6cm}
\begin{minipage}[t]{6.0cm}
\epsfig{figure=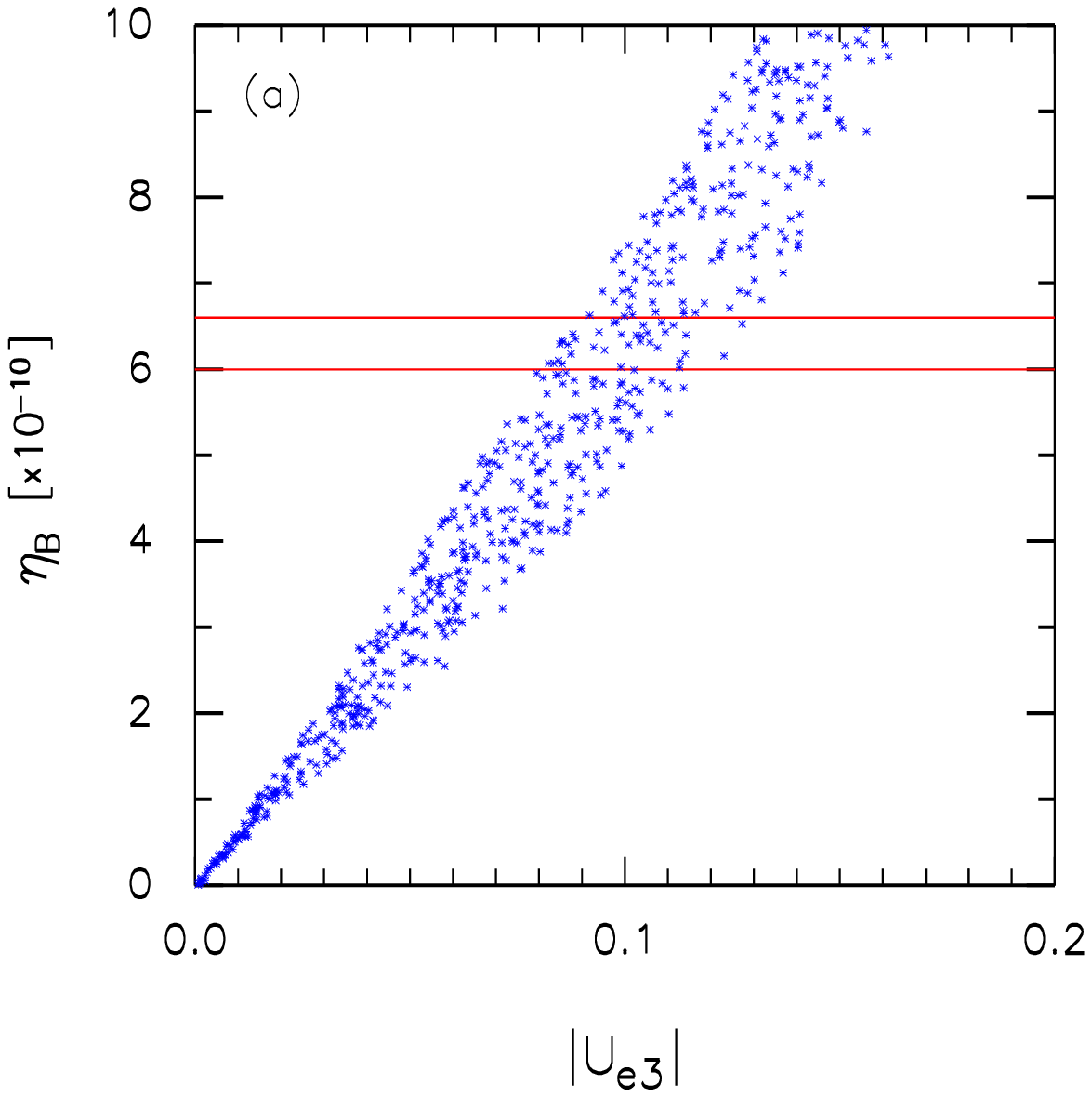,width=13cm,angle=0}
\end{minipage}
\hspace*{2.0cm}
\begin{minipage}[t]{6.0cm}
\epsfig{figure=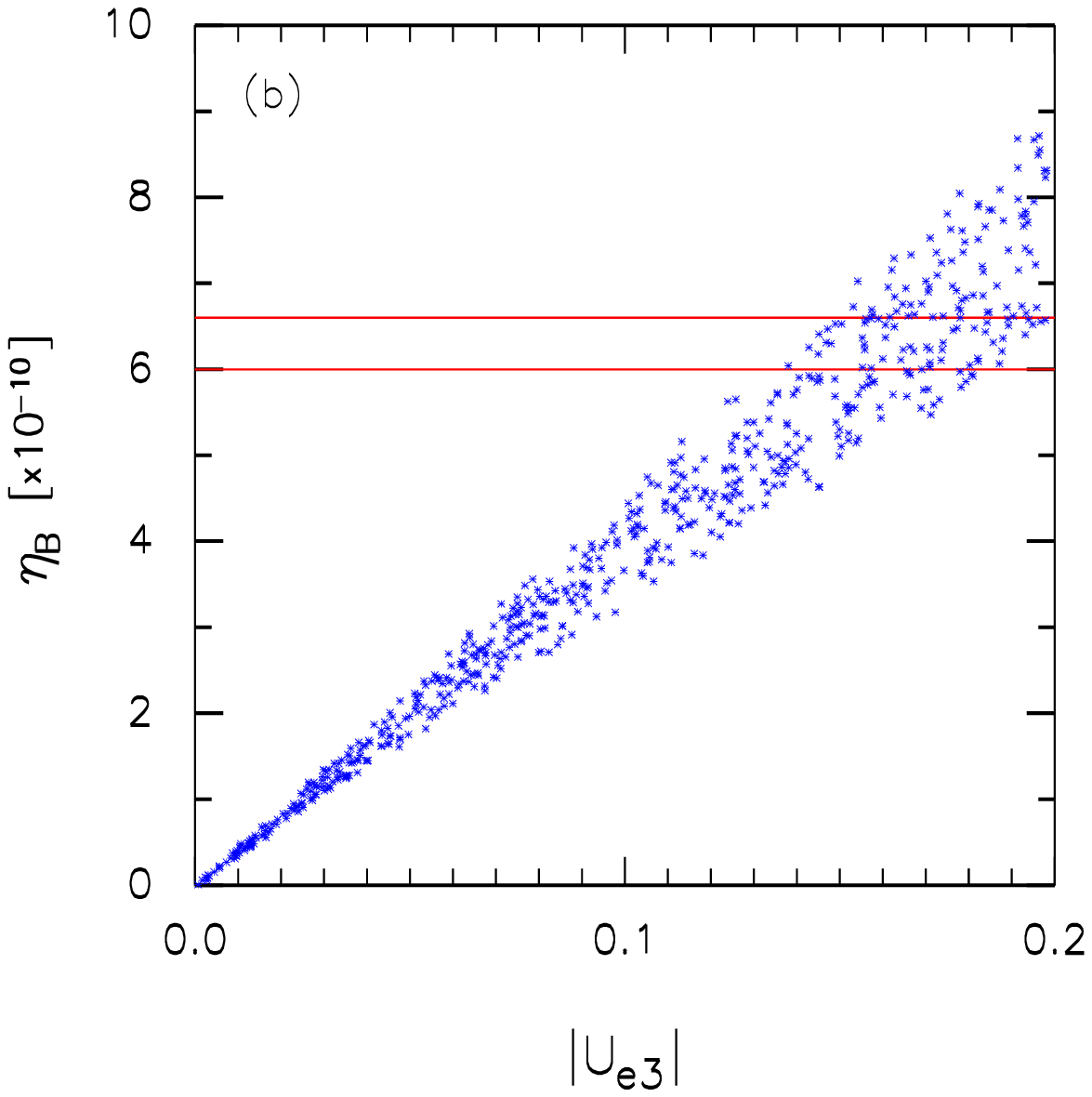,width=13cm,angle=0}
\end{minipage}
\vspace*{-5.5cm}
\caption{\label{Fig6}Predictions for the baryon asymmetry $\eta_B$
over $|U_{e3}|$ for (a) Case 2 and (b) Case 3 in Eq.~(\ref{cases}).
Here we take $\delta_N=10^{-6}$ and
$M_2=3\times 10^{6}$(GeV). The two horizontal lines are
the current bounds from the CMB observations \cite{cmb}.}
\end{figure}

In Fig.~\ref{Fig6}, we present the predictions for $\eta_B$ over
$|U_{e3}|$ for a sufficient degeneracy, $\delta_N=10^{-6}$, and
rather light $M_2=3\times10^{6}$ GeV. The two horizontal lines are
the current bounds from the CMB observations \cite{cmb},
$\eta_B^{\rm CMB}=(6.3\pm 0.3)\times 10^{-10}$ \cite{bari}. As
shown in Fig.~\ref{Fig6}, the current observation of $\eta_B^{\rm
CMB}$ can narrowly constrain the value of $|U_{e3}|$. Combining
the results presented in Fig.~\ref{Fig3} with those from
leptogenesis, we can pin down the CP phase $\alpha$, from which
the predictions of $m_{ee}$ and
 $\varepsilon$ are constrained, as can be seen from Fig.~\ref{Fig4}.
For example, if $|U_{e3}|$ is determined to be around $|U_{e3}| \sim 0.1$,
the CP phase $\alpha$ should be around $45^\circ$, which in turn leads to
$0.003~(0.0025) \leq m_{ee} \leq 0.008~(0.0035)$ for Case 2 (3).

\section{Remarks and Conclusions}

In order to achieve leptogenesis, we have demanded the breaking of degeneracy $M_2=M_3$
in heavy Majorana mass spectrum.
The lift of the degeneracy between $N_2$ and $N_3$ leads to the $\mu-\tau$ symmetry breaking
in the effective light neutrino mass matrix on top of the breaking due to nontrivial CP phase.
Although tiny breaking of the $\mu-\tau$ symmetry is enough for successful leptogenesis,
its breaking effect may affect our predictions for neutrino masses and mixing described in sec. III.
The most severe influence can be happened in the prediction of non-vanishing $\theta_{13}$.
So, let us estimate how much it  can be affected by  $\delta_N$ demanded for successful leptogenesis.
Taking $M_N={\rm Diag}[M_1,M_2,M_3]$ and
imposing $\alpha=0$ in Eq.~(\ref{meff}), we obtain
\begin{eqnarray}
\theta_{13}\simeq \frac{\omega(\kappa-1)}{\sqrt{2}(\omega^2-\frac{(\kappa-1)^2}{2})}
\frac{M_3-M_2}{M_3+M_2}.
\end{eqnarray}
Using the allowed regions of the parameters $\kappa$ and $\omega$ shown in
in Figs.~\ref{Fig1} and \ref{Fig2}, we estimate that
\begin{eqnarray}
\theta_{13} < 0.2\delta_N.
\end{eqnarray}
So, we find that the $\theta_{13}$ generated by the generic
$\mu-\tau$ symmetry breaking is less than $0.2^\circ$ even for
$\delta_N=10^{-2}$, which does not hurt the analysis for the
neutrino masses and mixing described in sec. III.
Although the mass splitting $\delta_N$ and the phase angle $\alpha$ are totally
independent in our consideration, both can generate the non-zero
$\theta_{13}$ as shown in Eqs. (21,40). However, as explained above
and in Fig. 4, the allowed ranges of the parameter space are quite
different, $i.e.$ $\theta_{13} < 13^\circ$ from $\alpha$ and
$\theta_{13} \sim 0^\circ$ from $\delta_N$.

Finding non-vanishing but small mixing element $U_{e3}$ would be
very important in the near future mainly because of completeness
of neutrino mixing and possible existence of leptonic CP
violation. Theoretically, non-vanishing $U_{e3}$ may be related
with a certain flavor symmetry breaking in the neutrino sector. In
this paper, we proposed a new scenario to break the $\mu-\tau$
symmetry so as to accommodate the non-vanishing $U_{e3}$ while
keeping maximal mixing for atmospheric neutrinos. Our scenario is
constructed in the context of a seesaw model, and the symmetry
breaking is achieved by introducing a CP phase in the Dirac Yukawa
matrix. Then, the resultant effective light neutrino mass matrix
generated through seesaw mechanism reflects the $\mu-\tau$
symmetry breaking which is parameterized in terms of the CP phase,
and the $\mu-\tau$ symmetry is recovered in the limit of vanishing
CP phase. We discussed how neutrino mixings and the neutrino
mass-squared differences can be predicted and showed how the
deviation of $\theta_{23}$ from the maximal mixing and
non-vanishing $U_{e3}$ depends on the CP phase in our scenario.

The prediction for the amplitude in neutrinoless double beta decay
has been also studied. However, the Dirac CP phase introduced
to break the $\mu-\tau$ symmetry does not lead to successful
leptogenesis. We found that a tiny breaking of the degeneracy
between two heavy Majorana neutrinos on top of the $\mu-\tau$
symmetry breaking through the CP phase can lead to successful
leptogenesis without much changing the results for the low energy
neutrino observables.  We also examined how leptogenesis can
successfully work out and be related with low energy neutrino
measurement in our scenario, and showed that our predictions for
the neutrino mixing can be severely constrained by the current
observation of baryon asymmetry. Future measurement for $U_{e3}$
would play an important role of test of our scenario and provide
us with some indication on baryon asymmetry in our Universe.
\\

\acknowledgments{
\noindent YHA was supported in part by Brain Korea 21 Program and
in part by  CHEP-SRC Program and
in part by the Korea Research Foundation Grant funded by the Korean Government
(MOEHRD)  No. KRF-2005-070-C00030.
CSK was supported
by the Korea Research Foundation Grant funded by the Korean Government
(MOEHRD) No. R02-2003-000-10050-0.
JL was supported in part by Brain Korea 21 Program and
in part by Grant No. F01-2004-000-10292-0 of KOSEF-NSFC International
Collaborative Research Grant.
SKK and JL were supported in part by the SRC program of KOSEF through
CQUeST with Grant No. R11-2005-021.
}

\end{document}